\documentclass[runningheads]{llncs}
\usepackage{graphicx}
\usepackage{comment}
\usepackage{amsmath,amsfonts,amssymb}
\usepackage{array,multirow,graphicx}
\usepackage{adjustbox}
\usepackage{booktabs}
\usepackage{tabu, colortbl}
\usepackage{xr}
\usepackage[dvipsnames]{xcolor}
\usepackage[super]{nth}
\usepackage{tikz}
\usepackage[colorlinks=true, allcolors=blue]{hyperref}
\usepackage[skip=10pt]{caption} 
\usepackage{subcaption}
\usetikzlibrary{svg.path} 
\usepackage{pgfplots} 
\usepackage{pgfplotstable} 

\definecolor{mycorrect}{rgb}{1, 0, 0} 
\definecolor{light_grey}{rgb}{0.6, 0.6, 0.6}

\begin{document}

\title{Towards Learning Contrast Kinetics 
with Multi-Condition Latent Diffusion Models}



\titlerunning{Towards Learning Contrast Kinetics}
%
%
%

%
\author{
Richard Osuala\inst{1,2,3}
\and
Daniel Lang\inst{2,3}
\and
Preeti Verma\inst{1}
\and
Smriti Joshi\inst{1} 
\and
Apostolia Tsirikoglou\inst{4} 
\and
Grzegorz Skorupko\inst{1} 
\and
Kaisar Kushibar\inst{1}
\and
Lidia Garrucho\inst{1}
\and
Walter H. L. Pinaya\inst{5}
\and
Oliver Diaz\inst{1,6}
\and
Julia Schnabel\inst{2,3,5}
\and
Karim Lekadir\inst{1,7}
}

\authorrunning{R. Osuala et al.}

\institute{Departament de Matemàtiques i Informàtica, Universitat de Barcelona, Spain
\email{richard.osuala@ub.edu}
\and
Helmholtz Center Munich, Munich, Germany
\and
Technical University of Munich, Munich, Germany
\and
Karolinska Institutet, Stockholm, Sweden
\and
Kings College London, London, UK
\and
Computer Vision Center, Bellaterra, Spain
\and
Institució Catalana de Recerca i Estudis Avançats (ICREA), Barcelona, Spain
}

\maketitle 

\begin{abstract}

Contrast agents in dynamic contrast enhanced magnetic resonance imaging allow to localize tumors and observe their contrast kinetics, which is essential for cancer characterization and respective treatment decision-making. However, contrast agent administration is not only associated with adverse health risks
, but also restricted for patients during pregnancy, and for those with kidney malfunction, or other adverse reactions
. With contrast uptake 
as key biomarker for lesion malignancy, 
cancer recurrence risk, and 
treatment response, it becomes pivotal to reduce the dependency on intravenous contrast agent administration
. To this end, we propose a multi-conditional latent diffusion model capable of acquisition time-conditioned image synthesis of 
DCE-MRI temporal sequences
. To evaluate medical image synthesis, we additionally propose and validate the Fréchet radiomics distance as an image quality measure based on biomarker variability between synthetic and real imaging data. Our results demonstrate our method's ability to generate realistic 
multi-sequence fat-saturated breast DCE-MRI 
and uncover the emerging potential of deep learning based contrast kinetics simulation. 
We publicly share our accessible codebase at \url{https://github.com/RichardObi/ccnet} and provide a user-friendly library for Fréchet radiomics distance calculation at \url{https://pypi.org/project/frd-score}.


\end{abstract}

\section{Introduction} 

Magnetic resonance imaging (MRI) and, in particular, dynamic contrast enhanced (DCE)-MRI are remarkably sensitive and effective modalities for tumor detection, localization and characterization and, thus, have become ubiquitous in clinical cancer treatment planning and monitoring. 
The uptake of contrast in DCE-MRI sequences
plays a pivotal role as biomarker for cancer detection, tumor molecular subtype and malignancy differentiation, as well as cancer recurrence and treatment response prediction \cite{wu2016breast,ground_truth}. 
However, intravenously-injected gadolinium-based contrast agents (CA) used in DCE-MRI have been associated with a wide range of concerns \cite{olchowy2017presence}, such as a risk of nephrogenic systemic fibrosis
, bioaccumulation in the brain, and its invasiveness causing a non-applicability in patient populations with pregnancy, adverse reactions, kidney malfunction or where consent is missing. Furthermore, due to its multiple temporal acquisitions, DCE-MRI is costly and time-consuming, prone to motion artifacts, and the contrast injection an uncomfortable procedure for patients \cite{zhang2023synthesis,osuala2023pre,olchowy2017presence}.

Public health institutions, such as the European Medicines Agency, recommend to restrict gadolinium-based CA \cite{euagency2017}, 
 which further emphasizes the need to develop alternative methods. A potentially faster, cost-effective, motion artifact-free, and non-invasive alternative is the 
synthetic generation of DCE-MRI using deep generative models. 
First studies applied generative adversarial networks (GANs) \cite{goodfellow2014generative} to generate post-contrast images from pre-contrast images \cite{muller2023using,zhang2023synthesis,osuala2023pre}. 
Recent works have further used diffusion models \cite{sohl2015deep} and latent diffusion models (LDMs) \cite{rombach2022high} to synthesize medical images 
\cite{pinaya2023generative,chambon2022adapting,khader2023denoising}, such as pre-contrast breast MRI conditioned on anatomical segmentation masks \cite{konz2024anatomically}. 
Despite their recent successes in computer vision, 
diffusion models and LDMs have, to the best of our knowledge, not been applied to pre- to post-contrast DCE-MRI translation.

A largely unaddressed aspect in medical image synthesis is domain-specific image quality evaluation. To date, popular methods, such as the Fréchet inception distance (FID) \cite{heusel2017gans}, 
are based on feature extractor models trained on natural image datasets. Despite a considerable domain gap, these methods are commonly applied to medical imaging data without alteration, thereby failing to capture medical nuances such as abnormalities \cite{chambon2022adapting}. Recently, an FID calculation based on a radiology domain-specific feature extractor was proposed \cite{osuala2023medigan}
, which nevertheless showed some volatility and no significant correlation with human judgement \cite{woodland2023importance}. A further limitation of such methods is the pretraining of the feature-extractor on 2D data (with unknown inherent biases), which is not applicable nor readily extendable to 3D medical images. 

In this work, we aim to address the aforementioned gaps, resulting in the following three contributions: 
\begin{itemize}
    \item Design,  implementation and validation of a multi-conditional latent diffusion model for pre- to post-contrast MRI synthesis.
    \item We present the first work that simulates 
    time-dependent contrast uptake on imaging data using diffusion models.
    \item We propose and validate the Fréchet radiomics distance (FRD), a novel radiology-specific quality evaluation method of 3D and 2D synthetic images based on biomarker variability.
\end{itemize}

\begin{figure} [ht]
\begin{center}
\includegraphics[height=5.75cm]{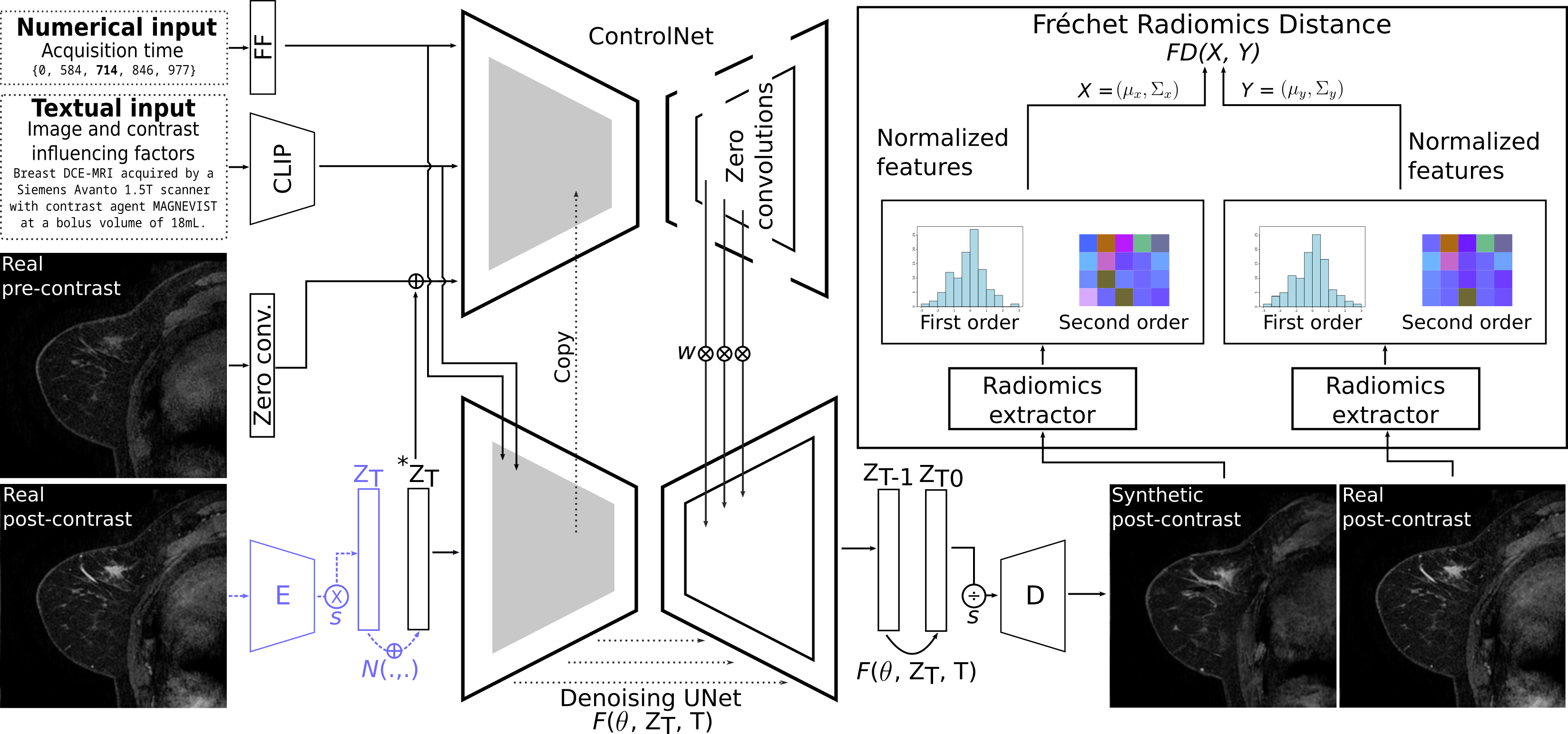}
\end{center}
\caption[] 
{\label{fig:overview} Overview of our proposed methods, including ContrastControlNet (CCNet) and the Fréchet radiomics distance (FRD). 
CCNet trains the denoising U-Net 
and the ControlNet in consecutive stages under contrast enhancement-specific conditioning (pre-contrast image, text, acquisition time). During inference, E is discarded (in violet) and, based on a random latent $z_T$ and $w$-weighted ControlNet guidance, the U-Net generates the post-contrast image latent $z_{T0}$. $z_{T0}$ is divided by factor $S$ and decoded via D into image space. Finally, FRD compares extracted real and synthetic imaging biomarker distributions.
}
\end{figure}

\section{Methodology}

\subsection{Multi-Condition Latent Diffusion Model}
Recently, diffusion models \cite{sohl2015deep,ho2020denoising} 
have emerged as a promising new class of generative models due to their exceptional ability to model complex distributions.
Such diffusion models consist of two processes, namely, a forward diffusion process and a reverse denoising process. The forward process is a Gaussian transition that gradually destroys the structure of a real data point $x_{0} \sim p(x_{0})$ by adding noise with different scales to obtain a series of noisy variables ${x_{1}, x_{2},..., x_{T}}$: 
 \begin{equation}
    q(\mathbf{x}_t|\mathbf{x}_0) = \mathcal{N}\left(\mathbf{x}_t| \sqrt{\bar{\alpha}_t}\mathbf{x}_0,(1-\bar{\alpha})\mathbf{I} \right).
\end{equation} 
The reverse process is parameterized by another Gaussian transition which gradually denoises $x_{T}$ in $T$ timesteps resulting in restoration of initial data point $x_{0}$:
\begin{equation}
p_\theta(\mathbf{x}_{t-1}|\mathbf{x}_{t})= \mathcal{N}\left(\mathbf{x}_{t-1}| \boldsymbol{\mu}_\theta(\mathbf{x}_t,t),\mathbf{\Sigma}_\theta(\boldsymbol{x}_t,t)\right).
\end{equation}
%
%
Latent diffusion models (LDM) \cite{rombach2022high} are designed to iteratively denoise a learned latent representation $z_{T}$ of image $x_{T}$ rather than operating directly on image space, which allows for more memory-efficient training and improved conditioning. The latter includes high-quality image synthesis based on textual input \cite{rombach2022high,chambon2022adapting}. However, text descriptions have been shown to be either inefficient or insufficient to accurately convey detailed controls upon the final generated images (e.g., to control the fine-grained semantic layout). 

ControlNet \cite{zhang2023adding} introduces 
plug-and-play condition encoders tailored for pre-trained diffusion models. We adopt ControlNet as auxiliary
encoder to integrate the conditioning pre-contrast image while preserving the integrity of the original post-contrast LDM generator. As shown in Fig. \ref{fig:overview}, the ControlNet encoder outputs are reintegrated into the diffusion model through zero-convolutional layers. We enhance this approach by extracting and propagating relevant textual metadata. We parse a clinical tabular dataset to assemble free-text prompts based on factors influencing 
DCE-MRI contrast manifestation. 
The final text is input to the denoising U-Net of both the LDM and the ControlNet via cross-attention and contains manufacturer, scanner, field strength, contrast agent, and bolus volume information. Furthermore, we integrate the time passed since pre-contrast acquisition into the model. Here, we extract pre- and post-contrast acquisition times from respective DICOM headers and input them as continuous variables into two dense layers before concatenating the resulting output with the timestep embedding of ControlNet and the denoising U-Net. Through optimization of the mean squared error between predicted and added noise per timestep, our ContrastControlNet (CC-Net) method learns to (i) extract meaningful patterns from pre-contrast, (ii) interpret conditions, (iii) reconstruct the corresponding post-contrast image, (iv) localize the lesion, and (v) predict realistic hyper- and hypo-intense lesion contrast uptake patterns.

\subsection{
Biomarker Variability as Image Quality Measure}

Despite the significant domain gap stemming from its feature extractor being trained on natural RGB images, FID \cite{heusel2017gans} is a commonly used metric in medical imaging, measuring the diversity and fidelity of synthetic images via real-synthetic distribution comparison. However, high-fidelity (and high-diversity) synthetic data can still be of low clinical utility and vice versa \cite{xing2023you}. For analyzing synthetic data, we note the need to ideally encompass (a) image utility indicators, (b) medical imaging domain-specific measurement, (c) scalability from 2D to 3D settings, and (d) explainable results based on interpretable features. Balancing these requirements, we propose measuring synthetic data quality based on imaging biomarker variability. In particular, we measure the feature distribution distance of multiple normalized biomarker values extracted from real and synthetic images. In radiology settings, 2D and 3D radiomics features can be extracted as non-invasive imaging biomarkers that quantify phenotypic characteristics \cite{lambin2012radiomics,van2017computational}. Further noting the capabilities of radiomics to capture 
tumor heterogeneity \cite{lambin2012radiomics}, or as predictor of treatment response \cite{ground_truth} and tumor subtype \cite{saha2018machine} in DCE-MRI, we introduce the Fréchet radiomics distance (FRD) as synthetic data quality measure.
While FRD feature inclusion choice is flexible, we compute FRD based on the common set of features suggested by \cite{van2017computational} including first-order statistics (n=19), 
co-occurrence gray level matrix (GLM) (n=24), run length GLM (n=16), size zone GLM (n=16), neighbouring gray tone difference matrix (n=5) and dependence GLM (n=14). Features are computed given the image and, optionally, a respective region of interest annotation.
%
As depicted in Fig. \ref{fig:overview}, to compute the FRD between two imaging datasets, we extract and normalize FRD features per image and dataset and model the resulting two feature sets as Gaussian distributions, allowing us to compute a distance between them.
Hence, for each image $x_{i}$ in a dataset, we extract a value $v_{ji}$ for each FRD feature $j$. Next, each $v_{ji}$ is 
min-max normalized based on the values $v_{j1}, v_{j2}, ..., v_{jn}$ over all images $X$ in the dataset. Next, the resulting values $v_{j}$ of feature $j$ are scaled to the common range observed for FID latent feature values, i.e. $[0, 7.456]$. This calibration later allows for interpretation of final FRD value and its comparison to FID, considering the intuition the image synthesis field has developed for FID value interpretation \cite{chambon2022adapting,osuala2023medigan,woodland2023importance,pinaya2023generative,xing2023you}.
The obtained synthetic and real feature sets $V$ are fitted to multivariate Gaussian distributions. These distributions are defined by their means ($\mu$) and covariance matrices ($\Sigma$). The FRD value is computed as the dissimilarity between real data $X$ and synthetic data $Y$ via the Fréchet distance defined as:
\begin{equation} \label{eq:FD}
    \begin{aligned}
        \textit{FD}(X, Y) = \lVert \mu_{X} - \mu_{Y} \rVert_{2}^{2} + \text{tr}(\Sigma_{X} + \Sigma_{Y} -2(\Sigma_{X}\Sigma_{Y})^{\frac{1}{2}}).
    \end{aligned}
\end{equation}

\section{Experiments and Results}

\subsection{Dataset and Implementation}
In this study, we use the public Duke-Breast-Cancer-MRI Dataset \cite{saha2018machine}.
The dataset and its imaging metadata 
encompasses 922 biopsy-confirmed breast cancer cases, each comprising one fat-saturated T1 sequence (pre-contrast) and up to 4 corresponding fat-saturated T1-weighted DCE sequences (post-contrast) with a median of 131 seconds passed between DCE sequences. 
The 1.5T or 3T MRI scans come in dimensions of either $320^2$, $448^2$ or $512^2$ in the coronal and sagittal planes, with varying slice numbers in the axial plane.
%
%
As we extract tumor-containing axial slices, we note considerable changes between pre- and post-contrast in non-tumor related areas (e.g., heart area). Hence, we crop the extracted slices first increasing the width and height of the tumor bounding box to half the width and height of the full image (e.g., $224^2$ in case of a $448^2$ image) thereby resulting in 
a tumor-containing single breast image. We split the dataset by patient into train (n=842), validation (n=50), and test (n=30) sets. All models were trained on a single GPU, either Nvidia A100 (80GB RAM) or RTX A6000 (48GB RAM), using Python3.11's pytorch, diffusers and monai-generative libraries \cite{pinaya2023generative}.


\subsection{Fréchet Radiomics Distance as Image Perturbation Correlate}
\begin{figure}[tb]
    \begin{subfigure}{.54\textwidth}
    \captionsetup{justification=centering}
    \caption[]{Image degradation scale example [in \%]}
      \includegraphics[width=\textwidth]{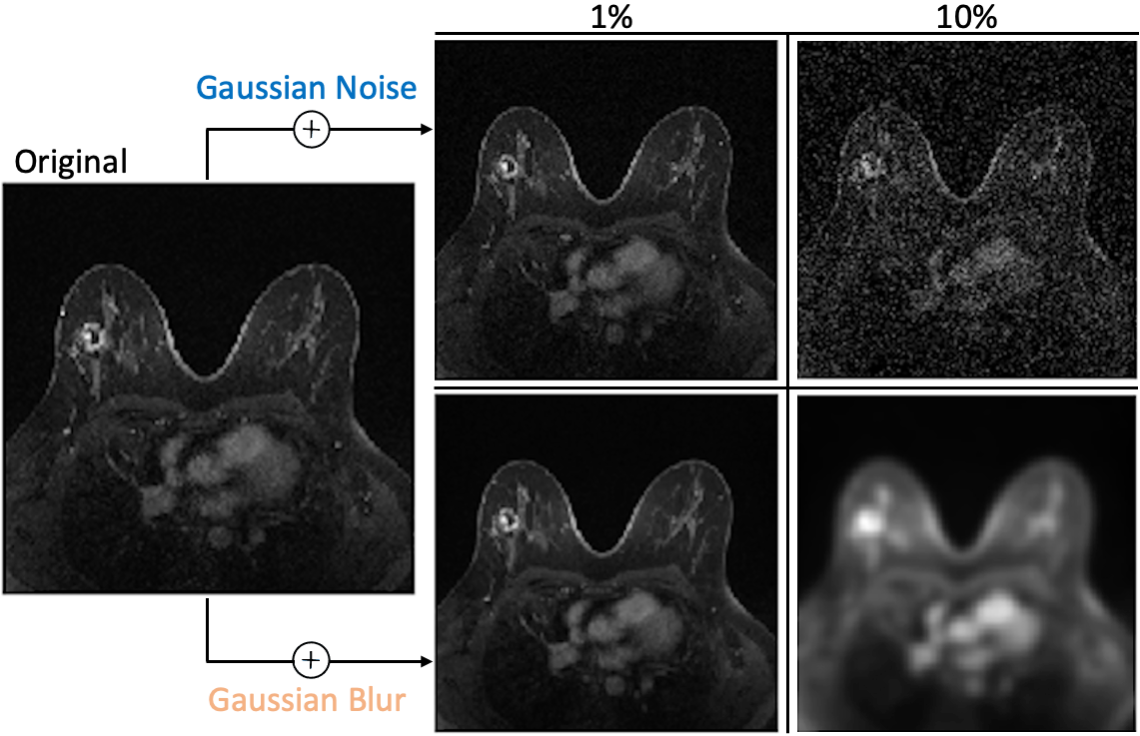}
      \label{fig:frd_example}
    \end{subfigure}
    \begin{subfigure}{.309\textwidth}
    \centering
    \caption[]{FRD [y] per \%-scale [x]} 
    \begin{tikzpicture}[scale=1.0,transform shape]
    \centering
    \begin{axis}[enlargelimits=0.05, 
            xmode=log,
            ymode=log,
            log ticks with fixed point,
            grid=both,
            scale only axis=true,
            width=1.0\textwidth, 
            height=1.00\textwidth, 
            ytick={0, 1, 5, 10, 50, 100, 250},
            xtick={0, 0.1, 1, 5, 10, 20, 50},
            ymin=0, 
            xmax=270.,
            ymin=0,
            xmax=50.,
            legend cell align=left,
            legend pos=south east,
            legend style={nodes={scale=0.75, transform shape}}
        ]
        \addplot+ [mark=*][
            scatter/classes={c={mark=*, mark options={line width=0.55pt}, draw=black, fill=blue}}, 
            scatter, mark=*, only marks, sharp plot,
            scatter src=explicit symbolic,
            nodes near coords*={\Label},
            visualization depends on={value \thisrow{label} \as \Label}
        ] table [meta=class] {
            x y class label
            1 0.717778 c \footnotesize{ABC}  
            5 9.992423 c \footnotesize{ABC} 
            10 36.634412 c \footnotesize{ABC}  
            20 109.500117 c \footnotesize{ABC}  
            50 212.479097 c \footnotesize{ABC}
        };
        \addplot+ [mark=*][
            scatter/classes={b={mark=diamond*, mark options={line width=0.55pt}, draw=black, fill=orange}}, 
            scatter, mark=*, only marks, sharp plot,
            scatter src=explicit symbolic,
            nodes near coords*={\Label},
            visualization depends on={value \thisrow{label} \as \Label}
        ] table [meta=class] {
            x y class label
            1 10.164644  b \footnotesize{ABC}  
            5 55.128728  b \footnotesize{ABC} 
            10 83.457771  b \footnotesize{ABC}  
            20 157.240380  b \footnotesize{ABC}  
            50 266.461072  b \footnotesize{ABC}
        };
        \addplot+ [mark=*][
            scatter/classes={e={mark=halfcircle*, mark options={line width=0.55pt}, draw=black, fill=blue}}, 
            scatter, mark=*, only marks, sharp plot,
            scatter src=explicit symbolic,
            nodes near coords*={\Label},
            visualization depends on={value \thisrow{label} \as \Label}
        ] table [meta=class] {
            x y class label
            1 0.42391 e \footnotesize{ABC}  
            5 11.63452 e \footnotesize{ABC} 
            10 40.3435 e \footnotesize{ABC}  
            20 80.9378 e \footnotesize{ABC}  
            50 145.8148 e \footnotesize{ABC}
        };
        \addplot+ [mark=*][
            scatter/classes={d={mark=halfdiamond*, mark options={line width=0.55pt}, draw=black, fill=orange}}, 
            scatter, mark=*, only marks, sharp plot,
            scatter src=explicit symbolic,
            nodes near coords*={\Label},
            visualization depends on={value \thisrow{label} \as \Label}
        ] table [meta=class] {
            x y class label
            1 9.60859  d \footnotesize{ABC}  
            5 25.28885  d \footnotesize{ABC} 
            10 43.77686  d \footnotesize{ABC}  
            20 86.833047  d \footnotesize{ABC}  
            50 164.184459  d \footnotesize{ABC}
        };
    \legend{3D Noise,3D Blur,2D Noise, 2D Blur}
    \end{axis}
    \end{tikzpicture} 
    \label{fig:frd_diagram}
    \end{subfigure}
\caption[]{(a) Image perturbation scales in breast MRI, (b) resulting Frechét radiomics distance (FRD) values (y-axis) per percentage scale (x-axis) per applied perturbation for 2D axial slices and for 3D volumes based on DCE-MRI post-contrast phase 1 data from 254 patient cases.

} 
\label{fig:FRD}
\end{figure}
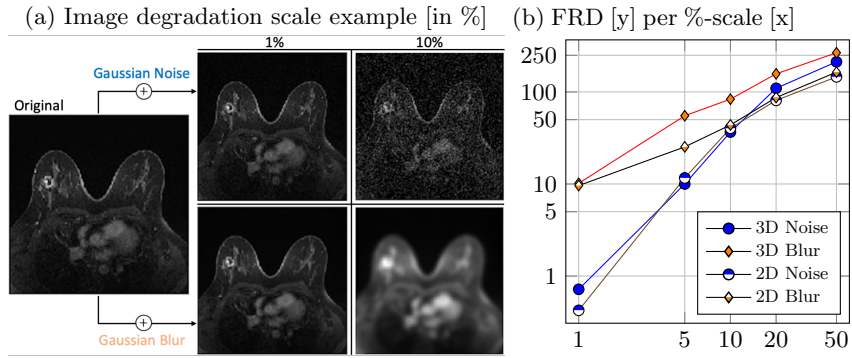
Adopting the validation strategy of the FID in its original publication \cite{heusel2017gans}, we observe the correlation between the Fréchet radiomics distance (FRD) and the amount by which the quality of an imaging dataset is reduced. Concretely, we compare FRD feature distributions between an unchanged imaging dataset and its quality-reduced equivalent, where the quality reduction is based on a scaling factor. As visualized in Fig. \ref{fig:frd_example}, we apply this procedure to 254 DCE-MRI (phase 1) cases of the Duke Dataset using Gaussian noise and Gaussian blurring for image perturbation on scales from 1\% up to 50\%. Unlike the FRD results shown in Table \ref{tab:pre_to_any_dce}, in Fig. \ref{fig:frd_diagram} we calculate the scores on full axial slices using the tumor mask. We observe
FRD monotonically increasing with 
perturbation scale demonstrating FRD's capability of 
capturing the quality-reduction level for both 2D and 3D data.

%
%
\subsection{Generation of DCE Sequences from Pre-Contrast Images} \label{sec:multi-experiments}
\begin{table}[ht!]
\centering
\caption{Synthetic image quality evaluation based on FRD, FID, LPIPS, and MSE metrics. Where applicable, results are reported with standard deviation and are based on 30 test cases consisting of 1010 images in each of the post-contrast phases P1, P2, and P3, and 417 images in phase P4. \textit{Any} refers to time-conditioned model training on all post-contrast phases. In \textit{+TXT} textual input is used in training and inference, \textit{+LDM} refers to LDM model training from scratch as opposed to stable diffusion \textit{2-1-base} fine-tuning. \textit{+CG} refers to a ControlNet guidance weight increase, e.g. from 1 to 1.6, during inference. \textit{+LT} stands for a 50 epoch longer training of ControlNet. Best results are in bold.
}
\resizebox{1.0\columnwidth}{!}{
\begin{tabular}{ll|cccc}
    \toprule
     \multicolumn{2}{c}{224x224 Single Breasts with Tumor} & \multicolumn{4}{c}{Metrics} \\
     \arrayrulecolor{black} \cmidrule(lr){0-1} \cmidrule(lr){3-6}
    Set 1 & Set 2 & FRD $\downarrow$ & FID $\downarrow$ & LPIPS $\downarrow$ & MSE $\downarrow$ \\
    \arrayrulecolor{black} \toprule
    Real Pre-Contrast & Real DCE-P1 & 49.07 & 68.20 & 0.223$\pm$.102 & 51.18$\pm$17.80 \\ 
    
    CC-Net$_{Any}$ P1 & Real DCE-P1 & 35.64 & \textbf{41.38}   & 0.192$\pm$.070 & 46.92$\pm$15.40\\

    
    CC-Net$_{Any+Txt}$ P1 & Real DCE-P1 & 39.98 & 42.85 & \textbf{0.186$\pm$.073} & 47.20$\pm$15.76 \\

    CC-Net$_{Any+Txt+LDM}$ P1 & Real DCE-P1 & 41.55 &  62.41 & 0.200$\pm$.074 & 49.17$\pm$14.48 \\    
    
    CC-Net$_{Any+Txt+LDM+CG1.6}$ P1 & Real DCE-P1 & \textbf{22.50} & 64.61  & 0.194$\pm$.072 & 46.12$\pm$14.21 \\

    CC-Net$_{Any+Txt+LDM+LT}$ P1 & Real DCE-P1 & 45.19 & 60.58 & 0.193$\pm$.073 & 48.08$\pm$14.44 \\    
    
    CC-Net$_{Any+Txt+LDM+CG1.6+LT}$ P1 & Real DCE-P1 & 37.70 & 62.21 & 0.192$\pm$.072 & \textbf{45.48$\pm$13.60}  \\
    
    \arrayrulecolor{light_grey} \cmidrule(lr){1-6}
    
    Real Pre-Contrast & Real DCE-P2 & 74.26  & 64.90 & 0.212$\pm$.095 & 50.00$\pm$16.85 \\
    
    CC-Net$_{Any}$ P2 & Real DCE-P2 & \textbf{58.07} & \textbf{40.36} & \textbf{0.191$\pm$.076} & \textbf{46.10$\pm$14.19} \\
        
    \arrayrulecolor{light_grey} \cmidrule(lr){1-6}
    
    Real Pre-Contrast & Real DCE-P3 & 84.13  & 60.96  & 0.208$\pm$.092 & 49.23$\pm$16.15 \\
    
    CC-Net$_{Any}$ P3 & Real DCE-P3 & \textbf{61.17} & \textbf{37.80} & \textbf{0.190$\pm$.074} & \textbf{45.75$\pm$13.74} \\

    \arrayrulecolor{light_grey} \cmidrule(lr){1-6}
    
    Real Pre-Contrast & Real DCE-P4 & 100.27 & 77.31 & 0.199$\pm$.078 & 52.48$\pm$12.96 \\
    
    CC-Net$_{Any}$ P4 & Real DCE-P4 &  \textbf{47.13} & \textbf{60.80} & \textbf{0.198$\pm$.075} & \textbf{50.36$\pm$14.26} \\

    
    
    
    
    %
    %
    \arrayrulecolor{black} \bottomrule
\end{tabular}
}
\label{tab:pre_to_any_dce}
\end{table}

%
%
\begin{figure}[tb]
\scalebox{1.0}{ 
\centering
    \begin{subfigure}{0.520\textwidth}
    \centering
    \caption[]{Qualitative test set results}
      \includegraphics[width=\textwidth]{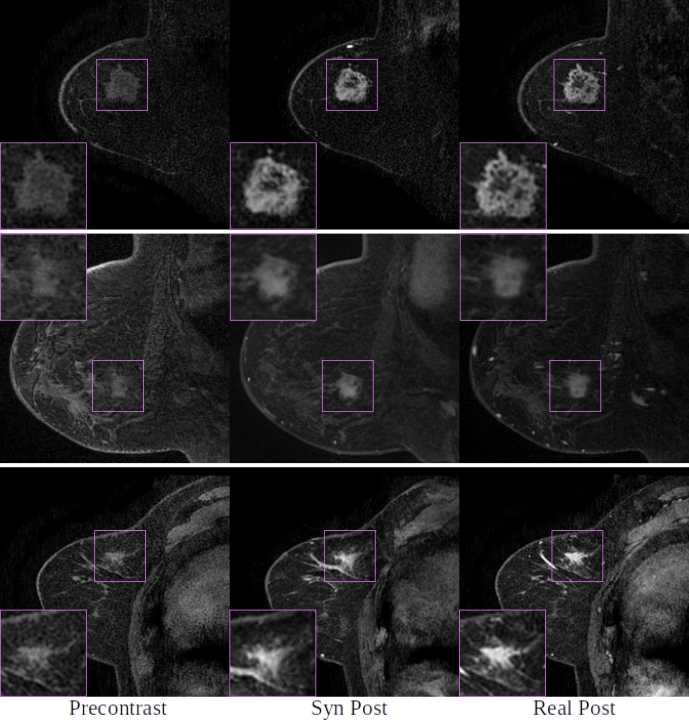}
      \label{fig:examples}
    \end{subfigure}
    \begin{subfigure}{0.4434\textwidth}
    \centering
     \caption[]{Tumor area contrast kinetics}
      \includegraphics[width=1.\textwidth]{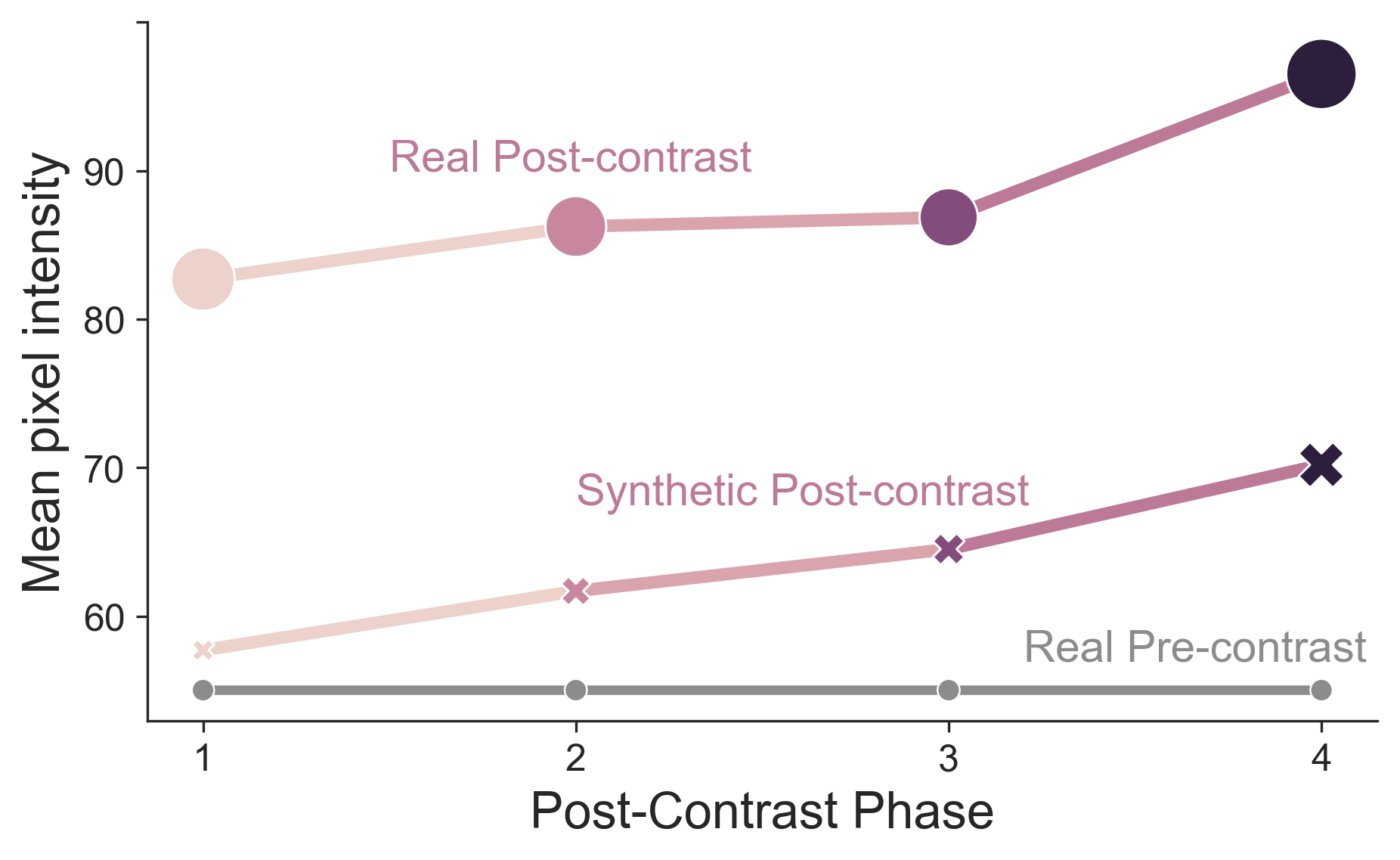}\quad
      \includegraphics[width=1.\textwidth]{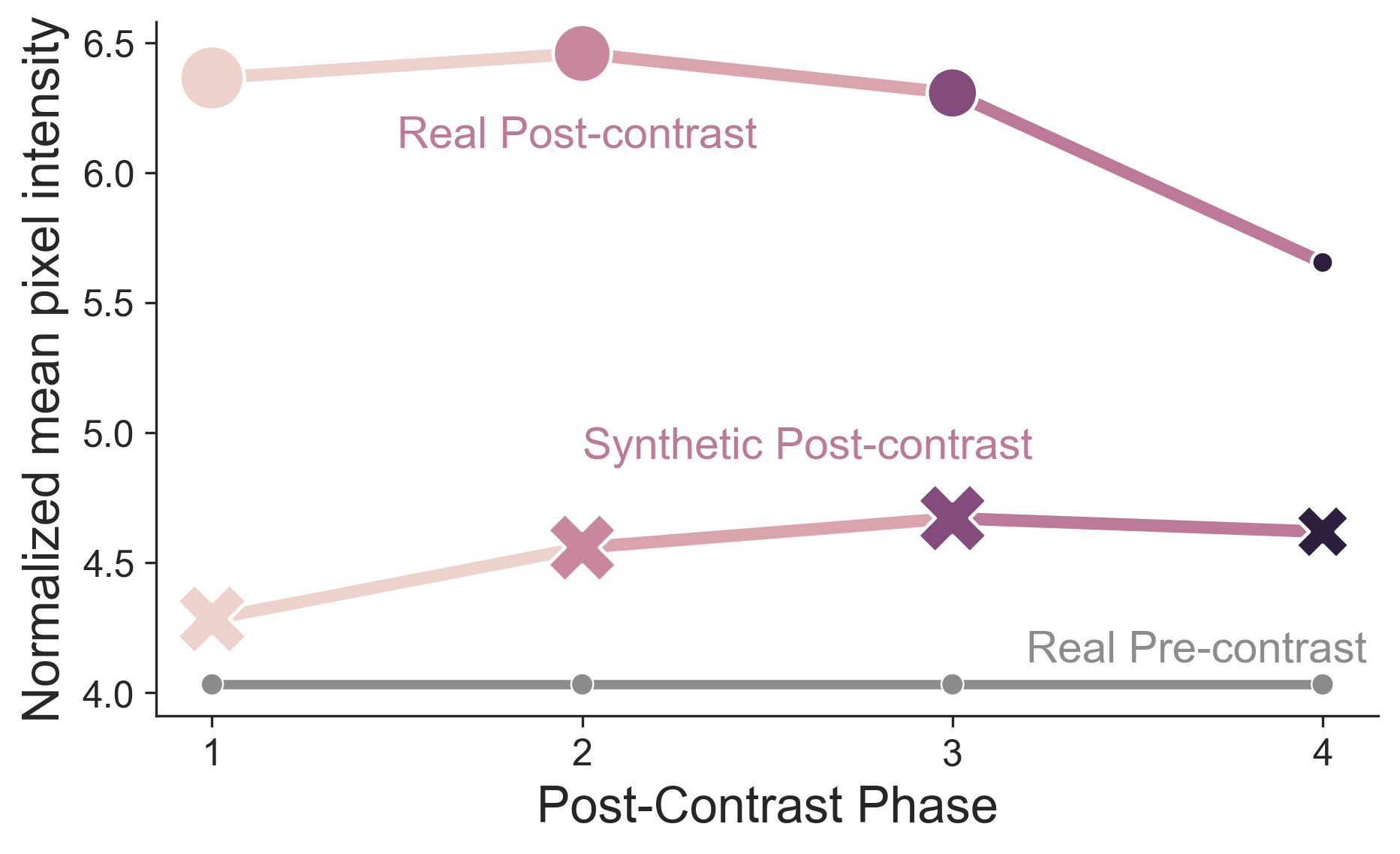}
      \label{fig:intensity_curves}
   
    \end{subfigure}
}
\caption[]{Qualitative (a) and quantitative (b) test set contrast enhancement. In (b), marker size represents the standard deviation of the mean intensity within the tumor region averaged across all test cases. When normalized, tumor region mean intensity is divided by mean intensity of the remaining tumor-free pixels.
}
\label{fig:results}

\end{figure}
In CC-Net, we first initialize a pretrained autoencoder (AE) from stable diffusion (SD), and then (b) proceed to train the denoising U-Net (i.e., LDM) before (c) training the ControlNet, after which we finally (d) run inference on the test set. 
In each step, we observe several hyperparameters to influence the empirical results. In (a), we note visible differences between the generally accurate breast MRI reconstructions of different pretrained SD AEs. We select the AE from SD \textit{2-1-base} over \textit{v1-5} and \textit{xl-base-1.0}. In (b), we notice a high dependence of output quality on scaling factor \textit{s} with which the AE representation is multiplied before being used in U-Net training. We find \textit{s}=0.1 to improve results upon the recommended \textit{s}=0.18215. We further identify a tendency for exploding gradients which was better addressed by clipping the gradient value rather than its norm (e.g., at 15). Both fine-tuning the SD U-Net, but also training it from scratch produced desirable outputs. We use a DDPM \cite{ho2020denoising}
noise scheduler with 1000 timesteps, AdamW as optimizer, and batch sizes of 8, 16, and 32, and learning rates of $2.5^{-5}$ and $5^{-5}$, for which we obtain similar results. 
The U-Net is trained for 100 epochs and selected from the epoch with lowest validation loss for further use in (c) and (d). In (c), we follow the hyperparameter setup from (b) with half the batch size and no gradient clipping.
In (d), we increase inference speed by using a DDIM \cite{song2021denoising} scheduler with 200 timesteps without visible performance decrease. We set text guidance scale to 1 as higher scales did not improve output quality, however, increasing the ControlNet guidance weight from 1 to a value in range (1, 1.6] enhanced perceived image quality.
In (a)-(c), following AE pretraining, the 224x224 input images are stacked in 3-channels and normalized in range [-1, 1] before applying small-scale intensity and affine augmentations. 
As observable in Fig. \ref{fig:examples}, the contrast-enhanced tumors account for a large part of the difference between pre- and post-contrast images. Bearing this in mind, we design experiments comparing the image and distribution-wise difference between our synthetic and the real post-contrast images against the difference between corresponding real pre-contrast and real post-contrast images. To quantify this difference, we use our FRD and the common FID \cite{heusel2017gans}, LPIPS \cite{zhang2018unreasonable} and mean squared error (MSE) metrics. Obtained results alongside ablations are summarized in Table \ref{tab:pre_to_any_dce}. Overall, CC-Net achieves substantially better results than the pre-contrast baseline across metrics and across post-contrast phases. Surprisingly, providing textual information not necessarily increases the performance of CC-Net$_{Any}$. 
Training the LDM from scratch instead of SD fine-tuning rather decreases performance (e.g., FID). The impact of longer ControlNet training is positive although marginal. Increasing the ControlNet guidance weight can be beneficial, and, in line with qualitative visual analysis, improves FRD and LPIPS.
In Fig. \ref{fig:intensity_curves}, using CC-Net$_{Any}$, we further analyze the pixel intensity distributions in the tumor area across phases aggregated over test cases. Synthetic images noticeably follow the principal trend of real contrast kinetics, despite a scale difference. A similar pattern is observable when taking contrast enhancements outside the tumor area into account by normalizing (dividing) mean tumor intensity by mean intensity of all other (tumor-free) pixels.

\section{Discussion and Conclusion}
We propose a multi-conditional latent diffusion model to translate pre-contrast into post-contrast images, thereby learning to highlight lesions by simulating their contrast uptake. We further condition the model on textual imaging metadata and continuous time passed since pre-contrast acquisition and demonstrate its synthesis capabilities on multi-sequence breast DCE-MRI data. We further contribute the Fréchet radiomics distance (FRD), a novel radiology-specific image quality metric measuring the distance between real and synthetic distributions of extracted interpretable imaging biomarkers. We validate FRD demonstrating its correlation with image perturbation scales on both 3D and 2D data. 
In future investigations, we aim to generate images of multiple DCE-MRI timepoints jointly 
and to map from the latent space of 3D autoencoders to the one from 2D-trained latent diffusion models.
In conclusion, our work paves the way for practical applications of deep generative models in MRI as a screening modality for unsupervised tumor detection and localization from pre-contrast MRI. It further constitutes a step towards improved treatment of patient populations where invasive contrast agent injection is contraindicated.

%
%

\subsubsection*{Acknowledgements.}
This study has received funding from the European Union’s Horizon 
research and innovation programme under grant agreement No 952103 (EuCanImage) and No 101057699 (RadioVal). It was further partially supported by the project FUTURE-ES (PID2021-126724OB-I00) and by grant FJC2021-047659-I from the Ministry of Science and Innovation of Spain. 
%
%

\bibliographystyle{splncs04}
\bibliography{temp.bib}



\end{document}